\documentclass[%
 reprint,
superscriptaddress,
 amsmath,amssymb,
 aps,
prb,
]{revtex4-2}

\usepackage[utf8]{inputenc}
\usepackage{graphicx}
\usepackage{dcolumn}
\usepackage{bm}
\usepackage{hyperref}
\hypersetup{colorlinks,linkcolor={blue},citecolor={blue},urlcolor={blue}}
\usepackage{braket}
\usepackage{booktabs,tabularx}
\usepackage{slashed}
\usepackage{dsfont}

\usepackage{todonotes,xcolor,tcolorbox}
\usepackage[normalem]{ulem}

\definecolor{dblue}{RGB}{31,119,180}
\definecolor{dblue_light}{HTML}{5AAAE3}
\definecolor{yellow1}{HTML}{E7EB90}
\definecolor{yellow2}{HTML}{FADF63}
\definecolor{yellow3}{HTML}{E6AF2E}
\definecolor{cite}{HTML}{2BB31E}
\definecolor{purple_light}{HTML}{DAB1DA}

\usepackage{relsize}

\renewcommand{\d}{{\rm d}}
\newcommand{\im}{{\rm i}}
\newcommand{\kb}{k_{\rm B}}
\newcommand{\e}{{\rm e}}

\newcommand{\fD}[1]{{#1}^{ \phantom{\dagger}}}

\newcommand{\bq}{{b {\bf q}}}

\newcommand{\ab}{{\alpha \beta}}

\newcommand{\aap}{{\alpha \alpha'}}
\newcommand{\bbp}{{\beta \beta'}}

\renewcommand{\vec}[1]{{\bf {#1}}}
\newcommand{\rr}{{\vec r}}
\renewcommand{\k}{{\vec k}}

\newcommand{\ki}{{\vec k}_{\rm i}}

\newcommand{\ks}{{\vec k}_{\rm s}}
\newcommand{\khs}{\hat{\k}_{\rm s}}
\newcommand{\khi}{\hat{\k}_{\rm i}}
\newcommand{\E}{{\vec E}}
\newcommand{\Ei}{{\E}_{\rm i}}

\newcommand{\Es}{{\E}_{\rm s}}
\newcommand{\q}{{\vec q}}

\renewcommand{\bq}{{\bar{\vec q}}}
\newcommand{\qs}{{\vec q s}}

\newcommand{\qsp}{{\vec q s'}}
\newcommand{\qpsp}{{\vec{q}' s'}}

\newcommand{\iqs}{{i \q s}}

\newcommand{\Rl}{{\vec R}_{L}}
\newcommand{\Rk}{{\vec R}_{K}}

\newcommand{\ei}{{\vec e}_{\rm i}}
\newcommand{\es}{{\vec e}_{\rm s}}
\newcommand{\eic}{{e}_{\rm i}}
\newcommand{\esc}{{e}_{\rm s}}

\newcommand{\keek}{{\ki ( \ei \es ) \ks}}

\newcommand{\w}{{\Omega}}
\newcommand{\wi}{{\omega_{\rm i}}}
\newcommand{\ws}{{\omega_{\rm s}}}

\newcommand{\dw}{{\ws - \wi}}
\newcommand{\dk}{{\ks - \ki}}

\newcommand{\Jqsp}{{J_{\q , ss'}}}
\newcommand{\Aqs}{{A_{\q  s}}}
\newcommand{\Aqsp}{{A_{\q  s'}}}
\newcommand{\Jqs}{{J_{\q s}}}
\newcommand{\viqs}{\fD{\vec{v}}_{\iqs}}

\newcommand{\crosssection}{\sigma}

\newcommand{\poangle}{{\theta}}
\newcommand{\crosssectionPara}{{\crosssection^\parallel_\poangle}}
\newcommand{\crosssectionPerp}{{\crosssection^\perp_\poangle}}

\newcommand{\ioio}{{(10\bar{1}0)}}
\newcommand{\oooi}{{(0001)}}
\newcommand{\ioi}{{(101)}}

\newcommand{\cDOS}{{\rm cDOS}}
\newcommand{\TPDOS}{{\rm 2PDOS}}
\newcommand{\TWDOS}{{2\omega{\rm -DOS}}}

\newcommand{\commaT}{{; T}}

\newcommand{\icm}{{cm$^{-1}$}}

\newcommand{\eiip}{{E$_2'$}}
\newcommand{\ai}{{A$_1$}}
\newcommand{\aito}{{A$_1$(TO)}}
\newcommand{\modeEi}{{E$_1$}}
\newcommand{\eito}{{E$_1$(TO)}}
\newcommand{\eii}{{E$_2$}}
\newcommand{\ailo}{{A$_1$(LO)}}
\newcommand{\eilo}{{E$_1$(LO)}}

\usepackage{glossaries}
\setacronymstyle{long-short}
\newacronym{gk}{GK}{Green-Kubo}
\newacronym{xc}{\emph{xc}}{exchange-correlation}
\newacronym{dft}{DFT}{density functional theory}
\newacronym{dfpt}{DFPT}{density functional perturbation theory}
\newacronym{md}{MD}{molecular dynamics}
\newacronym{aimd}{aiMD}{\emph{ab initio} molecular dynamics}
\newacronym{gan}{GaN}{gallium nitride}
\newacronym{bzs}{BaZrS$_3$}{barium zirconium sulfide}
\newcommand{\bzs}{\gls{bzs}}
\newacronym{po}{PO}{polarization oriented}
\newacronym{lda}{LDA}{local-density approximation}
\newacronym{gga}{GGA}{generalized gradient approximation}
\newacronym{tdep}{TDEP}{temperature-dependent effective potential}
\newacronym{dos}{DOS}{density of states}
\newacronym{jdos}{JDOS}{joint density of states}
\newacronym{2pdos}{2PDOS}{two-phonon density of states}
\newacronym{2wdos}{$2\omega$-DOS}{overtone density of states}
\newacronym{pbe}{PBE}{Perdew-Burke-Ernzerhof}
\newacronym{stdep}{sTDEP}{stochastic Temperature Dependent Effective Potential}
\newacronym{rmse}{RMSE}{root mean square error}
\newacronym{std}{STD}{standard deviation}
\newacronym{lo}{LO}{longitudinal optical}
\newacronym{to}{TO}{transverse optical}
\newacronym{sscha}{SSCHA}{stochastic self-consistent harmonic approximation}
\newacronym{tdscha}{TDSCHA}{time-dependent self-consistent harmonic approximation}

\newcommand{\pkg}[1]{\textsc{\color{darkgray}#1}}

\newcommand{\aims}{\pkg{{FHI-aims}}}
\newcommand{\qe}{\pkg{{Quantum Espresso}}}
\newcommand{\sokrates}{\pkg{SO3krates}}
\newcommand{\tdep}{\pkg{{TDEP}}}
\newcommand{\tdeptools}{\pkg{tdeptools}}
\newcommand{\ase}{\pkg{ASE}}

\begin{document}

\title{Ab initio theory of the non-resonant Raman effect in crystals at finite temperature in comparison to experiment: The examples of GaN and \texorpdfstring{BaZrS$_3$}{BaZrS3}}

\author{Florian Knoop}
\email{florian.knoop@liu.se}
\affiliation{Department of Physics, Chemistry and Biology (IFM), Link\"oping University, SE-581 83, Link\"oping, Sweden.}
\author{Nimrod Benshalom}
\affiliation{Department of Chemical and Biological Physics, Weizmann Institute of Science, Rehovot 76100, Israel}
\author{Matan Menahem}
\affiliation{Department of Chemical and Biological Physics, Weizmann Institute of Science, Rehovot 76100, Israel}
\author{Paul Gartner}
\affiliation{National Institute of Materials Physics, P.O.Box MG-7, Bucharest-Magurele, Romania}
\author{Tommaso Salzillo}
\affiliation{Department of Industrial Chemistry “Toso Montanari”, INSTM-UdR Bologna, Bologna 40129, Italy}
\author{Omer Yaffe}
\email{omer.yaffe@weizmann.ac.il}
\affiliation{Department of Chemical and Biological Physics, Weizmann Institute of Science, Rehovot 76100, Israel}
\author{Olle Hellman}
\email{olle.hellman@weizmann.ac.il}
\affiliation{Department of Physics, Chemistry and Biology (IFM), Link\"oping University, SE-581 83, Link\"oping, Sweden.}
\affiliation{Department of Molecular Chemistry and Material Science, Weizmann Institute of Science, Rehovot 76100, Israel}

\begin{abstract}
	We present an \emph{ab initio} theory of the non-resonant Raman scattering process in crystals at finite temperature in direct comparison with experiments.
	The theory naturally incorporates the scattering geometry and polarization dependence of the Raman process and the small but finite wave vectors of the phonons for correctly describing the scattering with longitudinal optical (LO) modes in optically anisotropic solids.
	We implement the theory for first-order Raman scattering and showcase the approach for wurtzite Gallium Nitride and the complex chalcogenide perovskite BaZrS$_3$ in comparison to experiment. We subsequently discuss several common estimates for second-order Raman scattering in complex materials,
	and highlight similarities and differences to established theoretical approaches and simulation protocols both from phonon theory and molecular dynamics.
\end{abstract}

\date{\today}
\maketitle
\glsdisablehyper

Raman spectroscopy is based on the Raman effect, which results from the dynamical modulation of the dielectric properties of a material or molecule through the atomic vibrations~\cite{Smekal.1923,RAMAN.1928}.
Compared to other forms of vibrational spectroscopy, such as neutron scattering, Raman spectroscopy is an accessible table-top technique available in many laboratories.
It is therefore an invaluable tool for the characterization of materials and molecules in a wide range of fields, and numerous variants such as stimulated or tip-enhanced Raman spectroscopy are in active use~\cite{Yu.2010,Das.2011,Jones.2019}.
While used primarily as a tool to characterize structural properties, recent years have led to a range of applications in the study of atomic vibrations  in catalysts~\cite{Hartman.2016,Hess.2021zm2,Loridant.2021}, molecular crystals~\cite{Putrino.2002,Pagliai.2008,Raimbault.2019}, ionic conductors~\cite{Julien.2018,Brenner.2020,Famprikis.2021}, perovskite photovoltaics~\cite{Ibaceta-Jana.2020xxc,Cohen.2022,Huang.2022}, and various semiconductors~\cite{Yu.2010}.

The comparison between experiment and atomistic simulations is particularly fruitful in elucidating the nuclear dynamics, as microscopic simulations allow us to directly assess and characterize the nuclear vibrations leading to scattering at a particular frequency.
Since the Raman effect originates in modulations of the electronic susceptibility, atomistic modeling requires an explicit treatment of electronic response, which can be obtained \emph{ab initio} in the framework of \gls{dft} and \gls{dfpt}~\cite{Baroni.1986qvh, Baroni.2001}.
On the other hand, the nuclear vibrations need to be described by solving the equations of motion for the nuclei. The most common approach for materials is to use the harmonic approximation to obtain the phonon eigenfrequencies and subsequently compute the modulation of the electronic susceptibility by the eigenmode displacements~\cite{Born.1954,Porezag.1996,Neugebauer.2002n7a,Lazzeri.2003,Veithen.2005,Skelton.2017}. This gives access to the Raman activity of a given mode and helps relate simulations with experiments at low temperatures, where anharmonic (phonon-phonon) interactions are typically weak, and the phonon spectral function is, therefore, sharply peaked around the eigenfrequencies.

At finite temperatures, anharmonic effects can lead to significant linewidth broadening (reduced phonon lifetime), phase transitions that change the Raman spectrum completely, and significant low-frequency scattering background from higher order phonon interactions~\cite{Menahem.2023}. In this regime, a realistic theoretical description of the Raman scattering process requires going beyond the harmonic approximation, either via direct solution of the equations of motion via \gls{md} simulations~\cite{Putrino.2002,Pagliai.2008,Thomas.2013sq4,Raimbault.2019,Ditler.2022,Hegner.2024}, or in the framework of lattice dynamics using anharmonic phonon theory such as perturbative or self-consistent approaches~\cite{Knoll.1995,Skelton.2017,Monacelli.2021xgu,Benshalom.2022,Miotto.2024,Hellman.2011, Errea.2013, Tadano.2015}.

While these approaches have been successfully employed in the past, particularly in the context of second-order Raman scattering, they typically rely on a simplified description of the Raman scattering process developed for cubic materials~\cite{Born.1947, Benshalom.2022}. This formalism does not account for the scattering geometry by assuming that the scattering wave vector $\q$ is strictly zero instead of small but finite, in the order of the wave vector of incident light in the optical range. This can lead to ambiguities when connecting the Raman spectrum to the vibrational dynamics, e.\,g., via response functions, and hinders theoretical understanding of anisotropic systems where the vibrational dynamics depend explicitly on the probing wave vector $\q$~\cite{Popov.2020}.

In this work, we revisit electrodynamic scattering theory and provide a comprehensive \emph{ab initio} derivation of the non-resonant \gls{po} Raman scattering process, including the scattering geometry \emph{without} assuming the harmonic approximation. The theory is valid for anisotropic crystals including direction-dependent LO/TO splitting, is fully compatible with \gls{dft} simulations, and we discuss how common expressions for the Raman scattering tensor follow as limiting cases.
By expanding the susceptibility in the nuclear displacements, we can reconnect this expression to phonon theory and enable temperature-dependent calculations of the spectral response in the framework of the \gls{tdep} method~\cite{Hellman.2011,Benshalom.2022,Knoop.2024}.

We demonstrate the approach for two anisotropic materials, wurtzite \gls{gan} used in light-emitting diodes and the complex chalcogenide perovskite \gls{bzs}, which is currently being investigated as an alternative to lead-based halide perovskites in optoelectronic applications~\cite{Yuan.2024rqs, Ye.2024, Sopiha.2022gzh, Choi.2022, Nishigaki.2020, Jaramillo.2019, Gross.2017, Wei.202094i, Kayastha.2023}.
For both materials, we achieve excellent agreement regarding mode assignment and polarization dependence, as well as the temperature dependence of spectral broadening and frequency shifts, besides typical systematic errors like underestimation of phonon frequencies~\citep{Benshalom.2022, Castellano.2023, Knoop.2024}.

The work is organized as follows:
In Sec.\,\ref{sec:theory}, we follow classical scattering theory to derive the Raman scattering process \emph{ab initio}, highlighting the role of small but finite scattering wave vectors $\q$ in the final expression. We furthermore discuss \gls{po} Raman highlighting the difference between unpolarized and isotropic averaging of the polarization orientations, before introducing first-order Raman scattering resulting from the first-order change of the electric susceptibility with the atomic displacement. The section is concluded by discussing approximate expressions for second-order Raman scattering in the harmonic approximation.
In Sec.\,\ref{sec:methods}, we discuss methodological and implementation details of the work, both for theoretical computations and experiment.
In Sec.\,\ref{sec:results}, we present and compare the experimental and theoretical results for \gls{gan} and \gls{bzs} obtained in this work, before concluding the work.

\section{Raman theory \label{sec:theory}}
\subsection{Direction-dependent spectral Raman tensor}
Following the development of classical scattering theory in Ref.~\citep{GRIFFIN.1968,Jackson.1998},
we describe the probing laser as an incident monochromatic electric field with amplitude $\Ei$ and frequency $\wi$ propagating in direction $\ki$
as
\begin{align}
	\Ei (\rr , t)
	= \Ei
	\e^{\im (\ki \cdot \rr - \wi t)} ~,
\end{align}
with wave vector
\begin{align}
	\ki =  n (\wi) \frac{\wi}{c} \khi~,
	\label{eq:ki}
\end{align}
where $n ( \wi)$ is the refractive index of the medium at frequency $\wi$, $c$ is the speed of light, and $\khi$ denotes the unit vector in the propagation direction.

Neglecting surfaces and boundaries, we label the unit cells of a crystal with $L$ and the Bravais lattice vectors pointing to them as $\Rl$. The induced dipole in a unit cell $L$ at time $t$ is given by
\begin{align}
	\vec p_{L} (t)
	= \chi (\Rl , t) \Ei \e^{\im ( \ki \cdot \Rl - \wi t)}~,
	\label{eq:p_iL(t)}
\end{align}
with $\chi (\Rl , t )$ the electronic susceptibility in cell $L$ modulated in time through the atomic motion -- this is the origin of the Raman effect.

We note that this expression assumes a \emph{local} and \emph{instantaneous} response and that the electronic susceptibility is independent of the incoming laser, i.e., for excitation frequencies well below the electronic band gap.

The susceptibility in a unit cell can be obtained from atomic contributions via averaging since typical Raman studies employ lasers in the optical range thus the wavelength of the field is $\lambda \simeq \mathcal O (100\,{\rm nm})$, much larger than the interatomic spacing. $\chi (\Rl)$ can, therefore, be obtained in the sense of a dipole approximation from averaging over a spatial extent $\ll 100$\,nm, i.e., one or more unit cells, as long as the result does not depend on the details of the averaging scheme. We comment on this point further in Sec.\,\ref{sec:chi.derivative}.

The frequency contribution of the oscillating dipole at scattering frequency $\ws$ is obtained via Fourier transformation,
\begin{align}
	\vec p_{L} (\ws)
	= \chi_{L} (\ws - \wi) \Ei \, \e^{\im  \ki  \cdot \Rl}~.
	\label{eq:pL(w)}
\end{align}
The accelerated dipole induces the scattered field at an observer point $\rr$ far away \citep{Jackson.1998},
\begin{align}
	\vec E_{{\rm s}, L} (\rr, \ws)
	= \frac{k_{\rm s}^2}{4 \pi \epsilon_0}
	\frac{\e^{\im k_{\rm s} r}}{r}
	\e^{- \im \ks \cdot \Rl}
	\khs \times \khs \times \vec p_L (\ws)~,
\end{align}
where we have approximated $\lvert \rr- \Rl \rvert \approx r - \khs \cdot \Rl$ to separate out the phase shift at $\Rl$, $\khs$ is the observer direction, and $\ks$ is the scattering wave vector defined analogously to Eq. \eqref{eq:ki}.

The total scattered field is obtained by summing over the individual contributions,
\begin{align}
	\Es (\rr, \ws)
	 & = \sum_L \vec E_{{\rm s}, L} (\rr, \ws) \\
	 & = \frac{k_s^2}{4 \pi \epsilon_0}
	\frac{\e^{\im k_s r}}{r}
	\khs \times \khs \times
	\chi (\dw, \dk) \Ei ~,
\end{align}
with the \emph{total crystal susceptibility}
\begin{align}
	\chi ( \dw, \dk)
	\equiv \sum_{L} \chi_{L} ( \dw )
	\,\e^{- \im [ \dk  ] \cdot \Rl }~.
	\label{eq:chi(w,k)}
\end{align}
We simplify the notation by introducing the \emph{Raman shift}
\begin{align}
	\w
	=
	\wi - \ws
	~,
	\label{eq:Raman.shift}
\end{align}
defined as \emph{positive} when the scattered frequency is \emph{smaller} than the incident frequency,
and the scattering wave vector
\begin{align}
	\q \equiv \ki - \ks~.
	\label{eq:q}
\end{align}
In this notation,
\begin{equation}
	\chi( \w, \q )
	\equiv
	\chi ( \wi - \ws, \ki - \ks)
	~,
\end{equation}
so that
\begin{equation}
	\chi( \ws - \wi, \ks - \ki)
	= \chi ( -\w, -\q )
	= \chi^\ast ( \w, \q )~.
\end{equation}

The differential scattering cross section $\crosssection$ for Raman is proportional to the ratio of the intensity of the scattered field in direction $\khs$ and polarization $\es \perp \khs$ to the incident field intensity in polarization $\ei$~\citep{Jackson.1998}.
It, therefore, depends on the scattering geometry with scattering vector $\q$ and the polarizations so that
\begin{align}
	\crosssection_{\ei \es} ( \w , \q  \commaT )
	 & \propto
	\ws^4
	\esc^\alpha \esc^\beta
	I^{\alpha \alpha', \beta \beta'} ( \w , \q  \commaT )
	\eic^{\alpha'} \eic^{\beta'}
	\label{eq:crosssection}
\end{align}
where superscript index denotes the vector component in the Cartesian coordinate $\alpha$ and $\beta$.
Non-essential prefactors have been omitted for clarity, and the \emph{spectral Raman tensor} is introduced,
\begin{align}
	I^{\alpha \alpha', \beta \beta'} ( \w  , \q \commaT )
	=
	\left\langle
	\chi^{\alpha \alpha'} ( \w  , \q ) \chi^{\ast, \beta \beta'} ( \w , \q )
	\right\rangle_T ~.
	\label{eq:Raman.tensor.frequency}
\end{align}
This tensor is typically proportional to the square of the susceptibility fluctuations \citep[p. 27]{Cardona.1982} and is temperature-dependent through the thermodynamic averaging $\langle \cdot \rangle_T$.

Using the susceptibility in Eq.\,\eqref{eq:chi(w,k)},
we get
\begin{align}
	 &
	I^{\alpha \alpha', \beta \beta'} ( \w , \q  \commaT )
	\\
	 & = \sum_{LK}
	\e^{- \im \q \cdot (\Rl - \Rk)}
	\left\langle
	\chi_L (\w )
	\chi_K^\ast (\w )
	\right\rangle_T
	\\
	 & =
	\sum_{LK}
	\e^{- \im \q \cdot (\Rl - \Rk)}
	\int \d t \, \e^{\im \w t}
	\left\langle
	\chi_L (t)
	\chi_K (0)
	\right\rangle_T
	~,
	\label{eq:Raman.rensor.FT.realspace}
\end{align}
via the Wiener-Khinchin theorem~\cite{Thomas.2013sq4}, where we note that $\chi_L (t)$ is real in the non-resonant setting. This relates the spectral Raman tensor given by the spectral density of the crystal susceptibility to the Fourier transform of the real-time and real-space susceptibilities.

The important difference to the form used by some of us in \citep{Benshalom.2022} is that the scattering geometry is still explicitly included via $\q = \ki - \ks$. Therefore, the phase relation between different unit cells in Eq.\,\eqref{eq:chi(w,k)} is preserved. Neglecting this phase relation by assuming $\q = \ki - \ks \approx 0 $ within the scattering region in Eq\,\eqref{eq:chi(w,k)} leads to the form
\begin{align}
	I^{\alpha \alpha', \beta \beta'} (\w \commaT )
	= \int \d t \, \e^{\im \w t}
	\left\langle
	\chi^{\alpha \alpha'} ( t ) \chi^{\beta \beta'} ( 0 )
	\right\rangle_T
	~,
\end{align}
in which the crystal is treated as one big molecule. This form is typically used in literature on second-order Raman effect in cubic crystals \citep{Born.1947, Cowley.1964, Benshalom.2022}, or molecular dynamics studies~\cite{Berger.2024bc8, Xu.2024, Rosander.2024, Berger.2024, Grumet.2024}, but it is not strictly valid in the usual experimental scenario when the penetration depth of the light spans several wavelengths in a single crystal.
Neglecting the photon momentum, therefore, does not allow to account for direction-dependent LO/TO splitting in optically anisotropic solids for small but finite scattering vectors $\lvert \q \vert \approx 2 k_{\rm i}$ in backscattering geometry, or the much smaller but still finite scattering vectors in small-angle forward scattering probing the Polariton region \citep{Henry.1965}.

In the case of birefringence, the discussion above needs to be adapted by accounting for the polarization dependence of the incident and scattered wave vectors $\ki$ and $\ks$, which will lead to a superposition of fields. This effect is particularly important in thin films where the size of the scattering region is constrained by the film thickness~\cite{Kranert.2015, Hildebrandt.2021}.

\subsection{Polarization oriented (PO) Raman}

As discussed in the previous section, the general Raman scattering cross section defined in Eq.\,\eqref{eq:crosssection} depends on the directions and polarizations of the incident and scattered fields, $\keek$, in Porto notation~\cite{Yu.2010}.

By denoting the angle with respect to the scattering plane spanned by $\khi$ and $\khs$, or some arbitrary reference in the case of backward or forward scattering ($\khi \parallel \khs$) as $\poangle$, the polarizations are unambiguously defined by $\khi$, $\khs$, and $\poangle$.
We then define the total \emph{unpolarized cross section} as
\begin{align}
	\sigma (\w )
	= \int \frac{\d \poangle}{2 \pi}
	\left(
	\crosssectionPara (\w ) + \crosssectionPerp (\w )
	\right)
	~,
	\label{eq:sigma.unpolarized}
\end{align}
where $\crosssectionPara$ denotes the \emph{parallel} configuration
\begin{align}
	\crosssectionPara (\w )
	 & \equiv \crosssection_{{\vec e}_{{\rm i}, \poangle} {\vec e}_{{\rm s}, \poangle}} (\w)
	~,
\end{align}
and $\crosssectionPerp (\ws )$ denotes the \emph{perpendicular} configuration with
\begin{align}
	\crosssectionPerp (\w)
	 & \equiv \crosssection_{{\vec e}_{{\rm i}, \poangle} {\vec e}_{{\rm s}, \poangle + \pi/2} } (\w)
	~,
\end{align}
where ${\vec e}_{{\rm i/s}, \poangle}$ are the polarization directions for a given angle $\poangle$.
In the special cases of backward or forward scattering with $\khi \parallel \khs$, we simply have ${\vec e}_{{\rm s}, \poangle} \parallel {\vec e}_{{\rm i}, \poangle}$ for parallel scattering, and ${\vec e}_{{\rm s}, \poangle + \pi/2} \perp {\vec e}_{{\rm i}, \poangle}$ for perpendicular scattering, respectively.

It is important to point out that the \emph{unpolarized} cross section defined in Eq.\,\eqref{eq:sigma.unpolarized} is not equivalent to the \emph{isotropic} average defined in Ref.~\cite{Porezag.1996,Neugebauer.2002n7a,Thomas.2013sq4,Skelton.2017}. In the present spectral framework, the isotropic average reads
\begin{align}
	\sigma^{\rm isotropic} (\w)
	=
	45 \alpha'^2 (\w)
	+
	7 \beta'^2 (\w)
	~,
	\label{eq:crosssection.iso}
\end{align}
with
\begin{align}
	\begin{split}
		9 \alpha'^2
		 & =
		\sigma_{xx} + \sigma_{yy} + \sigma_{zz}
		\\
		 &
		\quad + 2
		\left(
		\sqrt{ \sigma_{xx} \sigma_{yy}}
		+ \sqrt{ \sigma_{xx} \sigma_{zz}}
		+ \sqrt{ \sigma_{yy} \sigma_{zz}}
		\right)
		~,
	\end{split}
	\label{eq:crosssection.iso.alpha}
\end{align}
and
\begin{align}
	\begin{split}
		\beta'^2
		 & =
		\sigma_{xx} + \sigma_{yy} + \sigma_{zz}
		\\
		 &
		\quad -
		\left(
		\sqrt{ \sigma_{xx} \sigma_{yy}}
		+ \sqrt{ \sigma_{xx} \sigma_{zz}}
		+ \sqrt{ \sigma_{yy} \sigma_{zz}}
		\right)
		\\
		 &
		\quad + 3
		\left(
		\sigma_{xy} + \sigma_{xz} + \sigma_{yz}
		\right)
		~,
	\end{split}
	\label{eq:crosssection.iso.beta}
\end{align}
where $\sigma_{xx}, \sigma_{yy}, \ldots$ are the projections of the cross section defined in Eq.\,\eqref{eq:crosssection} on the Cartesian axes $\set{x, y, z}$, and all quantities depend on the frequency $\w$, omitted for readability in Eq.\,\eqref{eq:crosssection.iso.alpha} and \eqref{eq:crosssection.iso.beta}.
This isotropic average is valid for gas and liquid phases with random orientations of the microscopic scattering centers with respect to the laboratory frame. In a single crystal, however, the accessible components of the cross section are only those perpendicular to the scattering directions, and therefore only a subset of the components entering the isotropic cross section defined in Eq.\,\eqref{eq:crosssection.iso} will be measured in the unpolarized cross section defined in Eq.\,\eqref{eq:sigma.unpolarized} for a fixed scattering geometry.

A mixed case are polycrystalline solids: In this case, a full average over polarization directions and scattering direction needs to be performed while taking into account the explicitly $\q$ dependence of LO frequencies in the case of polar materials, see discussion in Sec.\,\ref{sec:loto} and Ref.\,\cite{Popov.2020}.

\subsection{First-order Raman}

To compute the first-order spectral Raman tensor via phonon theory, we expand the susceptibility to first order in the nuclear displacements $u_{iL}$,
\begin{align}
	\chi_L(t)
	= \sum_{i \gamma} \frac{\partial \chi_L(t)}{\partial u_{iL}^\gamma (t)} u_{iL}^\gamma (t)
	= \sum_{i \gamma} \chi_i^{ \gamma } u_{iL}^\gamma (t)~,
	\label{eq:chi.L(t)}
\end{align}
where we omit the second-rank Cartesian indices $\aap, \bbp$ on the tensor $\chi$ to simplify the notation.
Then Eq.\,\eqref{eq:Raman.rensor.FT.realspace} becomes
\begin{align}
	 &
	\langle
	\chi (\w, \q ) \chi^\ast (\w, \q )
	\rangle_T =
	\\
	 & \sum_{iL \gamma , jK \delta}
	\chi_i^{ \gamma} \chi_j^{ \delta}
	\e^{- \im \q \cdot (\Rl - \Rk)}
	\int \d t \, \e^{\im \w t}
	\langle
	u_{iL}^\gamma (t)
	u_{jK}^\delta (0)
	\rangle_T
	~,
\end{align}
where
$  \langle
	u_{iL}^\gamma (t)
	u_{jK}^\delta (0)
	\rangle
$
is the autocorrelation function for nuclear displacements, which is closely linked to the phonon propagator~\citep{GRIFFIN.1968}.
The displacements can be expanded using phonon amplitudes $\Aqs$ via~\cite{Benshalom.2022,Castellano.2023}
\begin{align}
	u_{iL}^\gamma (t)
	=
	\sum_{\qs}
	\text{v}^\gamma_\iqs \e^{\im \q \cdot \Rl} \Aqs (t)
	\label{eq:u.operator.expansion}
\end{align}
with lattice-periodic eigenvectors $\vec v_\iqs \e^{\im \q \cdot \Rl}$.
Then
\begin{align}
	 & \langle
	\chi (\w, \q ) \chi^\ast (\w, \q )
	\rangle
	_T
	\\
	 & =
	\sum_{ss'}
	\chi_{\bq s}
	\chi_{\q s'}
	\int \d t \, \e^{\im \w t}
	\langle
	A_{\bq s} (t)
	\Aqsp (0)
	\rangle
	_T
	\label{eq:chi(ws)}
	~,
\end{align}
with the first-order change of the susceptibility in mode space,
\begin{align}
	\chi_\qs
	= \sum_{i \gamma} \chi_i^\gamma \text{v}^\gamma_\iqs
	= \frac{\d \chi_L}{\d \Aqs}
	\equiv R_\qs
	~,
	\label{eq:chi.qs}
\end{align}
where $\bq \equiv - \q$, and where we identify $\chi_\qs$ with the second-rank Raman tensor commonly denoted as $R_{\q s}$ which determines the mode activity~\cite{Cardona.1982}.
The scattering wave vector $\q$ defined in Eq.\,\eqref{eq:q} is selected from the expansion in Eq.\,\eqref{eq:u.operator.expansion} via the closure relation of the lattice sums.

Treating the nuclei as quantum particles, we can use that \cite{GRIFFIN.1968, Benshalom.2022}
\begin{align}
	\int \d t \, \e^{\im \w t}
	\left\langle
	A_{\bq s} (t)
	\Aqsp (0)
	\right\rangle_T
	 & =
	( n ( \w \commaT ) + 1)
	\Jqsp ( \w )~,
	\label{eq:J}
	\\
	 & \equiv
	\Jqsp ( \w \commaT)
\end{align}
where $\Jqsp (\w)$ is the \emph{phonon spectral function}, $n (\w \commaT)$ is the Bose function at temperature $T$, and $\Jqsp (\w \commaT)$ is the Bose-weighted phonon spectral function. This result is equivalent to Eq.\,(10) in \cite{Benshalom.2022} for finite scattering $\q$.

A positive frequency $\w$ in the phonon spectral function, Eq.\eqref{eq:J}, is associated with phonon creation~\cite{Clerk.2010}.
However, from the light probe's perspective, phonon creation ($\w > 0$) means energy loss, i.e., photon annihilation with $\ws < \wi$.
By defining the Raman shift as $\w = \wi - \ws$ in Eq.\,\eqref{eq:Raman.shift}
we have, therefore, associated photon loss at frequency $\w$ with phonon creation, i.e., Stokes scattering.

For negative $\Omega$ with $\ws > \wi$, i.e., phonon absorption, we can use that the spectral function is odd in frequency~\cite{Baym.1961}, $\Jqsp (- \w) = - \Jqsp (\w)$, as well as $n(-\omega) = - n(\omega) - 1$~\cite{GRIFFIN.1968}, so that
\begin{align}
	\Jqsp ( - \w \commaT)
	 & =
	\exp \left( - \frac{ \hbar \w }{\kb T} \right)
	\Jqsp ( \w \commaT)
	~.
	\label{eq:J.negative.frequency}
\end{align}
This explains the ratio of scattering amplitudes between Stokes scattering ($\w > 0$) and Anti-Stokes scattering ($\w < 0$).
This result follows purely from a quantum mechanical treatment of the nuclear dynamics, while a classical treatment of the light scattering process is sufficient \cite{Knoll.1995}.

When the spectral function is diagonal in mode space, $\delta_{ss'}$, Eq.\,\eqref{eq:crosssection} reduces to
\begin{align}
	\crosssection_{\ei \es} (\w, \q  \commaT )
	\propto (\wi - \w)^4
	\sum_s \lvert \ei \cdot R_\qs \es \rvert^2 \Jqs (\w \commaT)
	~,
	\label{eq:crosssection.1}
\end{align}
which is the form typically used in the literature but with non-empirical inclusion of $\q$-dependence and an anharmonic spectral function~\cite{Popov.2020}.
In general, the spectral function does not need to be diagonal in mode space, as anharmonicity can lead to more complicated mode coupling. This can be particularly important in organic molecular crystals~\cite{Benshalom.2023}. In conventional inorganic crystals, this effect is often negligible.

\subsection{Second order Raman: Simple estimates \label{sec:theory.secondorder.raman}}

Following earlier work in Ref.\,\cite{Benshalom.2022,Menahem.2023}, the second-order spectrum can be obtained by an expansion of the susceptibility to second-order in the nuclear displacements~\cite{Born.1947}, leading to the convolution
\begin{align}
	\mathcal{S}_2(\w  \commaT)
	\propto &
	\sum_{\q, ss'} \chi''_{\q , ss'}
	J_{\q s} (\w  \commaT)
	\ast
	J_{\bq s'} (\w  \commaT)
	~,
	\label{eq:S2.1}
\end{align}
where $J_{\q s} (\w  \commaT)$
is the Bose-weighted spectral function as defined earlier, for which $J_{\q s} (\w  \commaT) = J_{\bq s} (\w  \commaT)$.
In the harmonic approximation, we have
\begin{align}
	J_{\q s} (\w)
	 & =
	- \delta (\w  + \omega_\qs)
	+ \delta (\w  - \omega_\qs)
	~,
	\quad \text{and}
	\\
	J_{\q s} (\w  \commaT)
	 & = n_{\q s} \delta (\w  + \omega_\qs)
	+ (n_\qs + 1) \delta (\w - \omega_\qs)
	,
	\label{eq:J_qs.bose.harmonic}
\end{align}
and therefore~\cite{Benshalom.2022}
\begin{widetext}
	\begin{align}
		\begin{split}
			J_\qs (\w  \commaT)  \ast  J_\qsp (\w  \commaT)
			= &
			[n(\w) + 1 ]
			(1 + n_\qs + n_\qsp)
			\left[ \delta (\w + \omega_\qs + \omega_\qsp) - \delta (\w - \omega_\qs - \omega_\qsp) \right]
			\\
			  &
			~~+
			[n(\w) + 1 ]
			(n_\qs - n_\qsp)
			\left[\delta (\w - \omega_\qs + \omega_\qsp) - \delta (\w + \omega_\qs - \omega_\qsp) \right]
			~.
			\label{eq:J_q*J_q.2}
		\end{split}
	\end{align}
\end{widetext}

Several simplified expressions for second-order Raman scattering can be inferred from here.

\subsubsection{2PDOS}
By setting $\chi''_{\q, ss'} = 1$ in Eq.\,\eqref{eq:S2.1}, we get
\begin{align}
	\mathcal{S}_2^\TPDOS{} (\w \commaT)
	 & \propto
	\sum_{\q, ss'}
	J_\qs (\w \commaT)  \ast  J_\qsp (\w \commaT)
	\\
	 & =
	g_2^{(1)} (\w \commaT)
	+
	g_2^{(2)} (\w \commaT)
\end{align}
with the Bose-weighted \gls{jdos} functions~\cite{Cardona.1982,Okubo.1983,Togo.2022}
\begin{widetext}
	\begin{align}
		g_2^{(1)} (\w \commaT)
		 & =
		[n(\w) + 1 ]
		\sum_{\q, ss'}
		(1 + n_\qs + n_\qsp)
		\left[ \delta (\w + \omega_\qs + \omega_\qsp) - \delta (\w - \omega_\qs - \omega_\qsp) \right]
		\\
		g_2^{(2)} (\w \commaT)
		 & =
		[n(\w) + 1 ]
		\sum_{\q, ss'}
		(n_\qs - n_\qsp)
		\left[\delta (\w - \omega_\qs + \omega_\qsp) - \delta (\w + \omega_\qs - \omega_\qsp) \right]~,
	\end{align}
\end{widetext}
which describe crystal-momentum-conserving $(\q + \q' = 0)$ phonon recombination and decay. These two functions together give the Bose-weighted \gls{2pdos},
\begin{align}
	g_2 ( \w  \commaT )
	 & =
	g_2^{(1)} ( \w  \commaT ) + g_2^{(2)} ( \w  \commaT )
	~,
	\label{eq:2pdos}
\end{align}
which is sometimes used to estimate second-order Raman scattering~\cite{Cardona.1982,Strauch.1996,Hildebrandt.2023byp}.

\subsubsection{\texorpdfstring{$2\omega$-DOS}{2w-DOS}}
An even simpler estimate can be obtained by setting $\chi''_{\q, ss'} = \delta_{ss'}$ in Eq.\,\eqref{eq:S2.1}, which yields the Bose-weighted \gls{2wdos}~\cite{Weinstein.1972,Weber.1976,1975am,Cardona.1982,Merlin.2000,Baroni.2001},
\begin{align}
	 & \mathcal{S}_2^{\TWDOS}  (\w \commaT)
	\\
	 & \quad \propto
	\sum_{\q s}
	J_\qs (\w \commaT)  \ast  J_\qs (\w \commaT)
	\\
	 & \quad =
	\sum_{\qs}
	n_\qs^2
	\delta ( \w + 2 \omega_\qs )
	+
	( n_\qs + 1 ) ^2
	\delta ( \w - 2 \omega_\qs )
	\\
	 & \quad \equiv
	g_{2 \omega} ( \w \commaT)
	~.
\end{align}
Estimates based on the \gls{2wdos} have been used for a range of systems inlcuding Si~\cite{Temple.1973,Wang.1973}, diamond~\cite{Windl.1993}, GaSb~\cite{Sekine.1976}, SiC~\cite{Windl.1994}, Ga$_2$O$_3$~\cite{Janzen.2022}, and CuI~\cite{Hildebrandt.2023byp}.

\subsubsection{DOS convolution}
As a third estimate, we relax the requirement of crystal-momentum conservation and estimate~\cite{Menahem.2023}
\begin{align}
	\mathcal{S}_2^{\cDOS}(\w \commaT)
	 & \propto
	\sum_{\q, ss'}
	J_\qs (\w \commaT)  \ast  J_\qsp (\w \commaT)
	\\
	 & \approx
	\sum_{\q \q', ss'}
	J_\qs (\w \commaT)  \ast  J_\qpsp (\w \commaT)
	\\
	 & =
	g( \w \commaT ) \ast g (\w \commaT)
	\label{eq:g*g.1}
	~,
\end{align}
where $g (\w \commaT )$ is the ordinary Bose-weighted one-phonon \gls{dos},
\begin{align}
	g ( \w \commaT )
	=
	\sum_{\qs}
	n_{\q s} \delta (\omega + \omega_\qs)
	+ (n_\qs + 1) \delta (\omega - \omega_\qs)
	~.
\end{align}
We denote the \gls{dos} autoconvolution in Eq.\,\eqref{eq:g*g.1} as {\cDOS}.

In summary, we have
\begin{align}
	\mathcal{S}_2^{\TPDOS} (\w \commaT)
	 & =
	g_2 (\w \commaT)
	\label{eq:S.2PDOS}
	\\
	\mathcal{S}_2^{\TWDOS} (\w \commaT)
	 & =
	g_{2 \omega} (\w \commaT)
	\label{eq:S.2wDOS}
	\\
	\mathcal{S}_2^{\cDOS} (\w \commaT)
	 & =
	g( \w \commaT ) \ast g (\w \commaT)~.
	\label{eq:S.cDOS}
\end{align}
We discuss the relevance of these estimates for complex materials further below.


\section{Methods \label{sec:methods}}

\subsection{Stochastic Temperature Dependent Effective Potential (sTDEP) \label{sec:stdep}}

While the basic theory for lattice dynamics in the framework of \gls{tdep} has been presented elsewhere~\cite{Hellman.2011, Hellman.2013, Benshalom.2022, Castellano.2023}, we briefly discuss essential aspects for completeness.

We use the \gls{stdep} scheme to compute phonons described by the force constants $\{ \Phi_{ij}^{\ab} (T) \}$ at a temperature $T$~\cite{Shulumba.2017}. This scheme is formally equivalent to the QSCAILD method described in Ref.~\cite{Roekeghem.2021} and relies on minimizing the mean-square error of the harmonic model forces,
\begin{align}
	S [\Phi, T]
	 & \equiv
	\sum_{i \alpha} \left\langle ( f_i^\alpha - f_{\Phi; i}^\alpha )^2 \right\rangle_{\Phi, T}~,
	\label{sm.eq:S1}
\end{align}\
where $\langle \cdot \rangle_{\Phi, T}$ denotes a thermal average with respect to the force constants $\Phi$ at temperature $T$, $ij , \alpha\beta$ denote atomic and Cartesian indices. The harmonic force $f_{\Phi; i}^\alpha$ is given by $f_{\Phi; i}^\alpha = - \sum_{j \beta } \Phi_{ij}^\ab  u_j^\beta$, where $\vec u_j$ are the atomic displacements from the thermal average positions, which are obtained by minimizing the average forces $\langle f_i^\alpha \rangle_{\Phi, T}$, respectively.

As shown and discussed in Ref.~\cite{Roekeghem.2021, Bianco.2017}, the least-squares solution to Eq.~\eqref{sm.eq:S1} is given by
\begin{align}
	\Phi_{ij}^\ab  (T)
	= - \left\langle u_i^\alpha u_j^\beta \right\rangle_{ \Phi, T}^{-1} \left\langle f_i^\alpha u_j^\beta \right\rangle_{ \Phi, T}^{\phantom{-1}}
	\equiv
	\left\langle \frac{\partial^2 V}{\partial u_i^\alpha \partial u_j^\beta} \right\rangle_{\Phi, T}~,
	\label{eq:phi.stdep}
\end{align}
\textit{i.e.}, the configuration average of the second derivative of the potential energy $V$ with respect to the effective harmonic distribution of atomic positions at target temperature $T$. In \gls{stdep}, the solution is obtained self-consistently by starting from a trial set of force constants~\cite{Shulumba.2017}, and solving Eq.\,\eqref{eq:phi.stdep} until convergence. This solution is equivalent to minimizing the Gibbs-Bogoliubov free energy and yields the best trial free energy of harmonic form similar to the \gls{sscha} method~\cite{Errea.2013,Bianco.2017,Monacelli.2021}. Further practical details are given in \cite{Benshalom.2022, Knoop.2024, Shulumba.2017,Castellano.2023qbw}.

After second-order force constants are converged, third-order force constants are fitted to minimize the residual force~\cite{Hellman.2013oi5}. From these, phonon lifetimes are computed as described in \cite{Hellman.2013oi5}, and the phonon spectral functions are computed as described in \cite{Li.2014,Romero.2015,Shulumba.20179s8e,Benshalom.2022}. While beyond the scope of the current work, we note that including fourth-order force constants is possible as well~\cite{Klarbring.2020vk,Wang.2023us,Castellano.2024}.

\subsection{Lattice dynamics for polar materials \label{sec:loto}}
To describe long-range electrostatic interactions in polar materials, it is common practice to use the \emph{ansatz} popularized by Gonze and coworkers for the interaction of atomic dipoles $\vec p_i = Z_i \vec u_i$ induced by nuclear displacements $\vec u_i$, where $Z_i$ is the Born effective charge tensor of atom $i$~\cite{Gonze.1994, Gonze.1997}:
In an anisotropic dielectric, the dipole-dipole force constants can be written as~\cite{Gonze.1997}
\begin{align}
	\left.\Phi^{\mathrm{dd}}\right|_{i j} ^{\alpha \beta}
	 & =\sum_{\gamma \delta} Z_i^{\alpha \gamma} Z_j^{\beta \delta} \widetilde{\Phi}_{i j}^{\gamma \delta}
	~,
	\label{eq:phi.polar.1}
	\\
	\tilde{\Phi}^{\alpha \beta}_{ij}
	 & =(\operatorname{det} \epsilon)^{-1 / 2}
	\left[\frac{\left(
			\epsilon^{-1}\right)^\ab
		}{
			\lvert \Delta_{ij} \rvert_\epsilon^3}
		-3 \frac{\Delta_{ij}^\alpha \Delta_{ij}^\beta}{\lvert \Delta_{ij} \rvert_\epsilon^5}\right]
	~,
	\label{eq:phi.polar.2}
\end{align}
where $\epsilon = 1 + 4 \pi \chi$ is the ion-clamped dielectric tensor of the medium which can be used to define a metric space in which $\Delta_{ij}^\alpha = \sum_\beta (\epsilon^{-1})^\ab r_{ij}^\beta$ is the displacement vector between atom $i$ and $j$, and
$\lvert \Delta_{ij} \rvert_\epsilon = \sqrt{{\bf \Delta}_{ij} \cdot {\bf r}_{ij}}$~\cite{Landau.2013}.
Fourier transforming this expression into reciprocal space yields the well-known non-analytical dynamical matrix in the $\q \to 0$ limit~\cite{Cochran.1962,Gonze.1994},
\begin{align}
	\left. C^{\rm NA} \right|^{\ab}_{i j}(\mathbf{q} \rightarrow 0)
	=
	\frac{4 \pi}{V_{\rm uc}}
	\sum_{\alpha' , \beta'}
	\frac{\left(q^{\alpha'} Z_{i}^{\alpha' \alpha}\right)\left(q^{\beta'} Z_{j}^{\beta' \beta}\right)}{\mathbf{q} \cdot \epsilon \mathbf{q}}
	~,
	\label{eq:dyn.mat.na}
\end{align}
where $V_{\rm uc}$ is the unit cell volume. This mechanism is responsible for LO/TO splitting in polar materials, where an extended \gls{lo} mode generates a macroscopic electric field, and this extra work results in a higher frequency compared to the \gls{to} counterpart~\cite{Lyddane.1941, Cochran.1962}.

As comprehensively discussed by Zhou and coworkers in Ref.\,\cite{Zhou.20197fs}, the force constants defined in Eq.\,\eqref{eq:phi.polar.2} do not automatically fulfill basic physical properties like hermiticity and the acoustic sum rule in low-symmetry materials~\cite{Pick.1969}. Several workarounds have been presented in the literature, such as imposing hermiticity on the full force constants via modifying the short-range force constants~\cite{Zhou.20197fs}, or by brute-force symmetrization~\cite{Lin.2022}. In the present work, we implement an approach similar to the one suggested in~\cite{Zhou.20197fs} by modifying the short-range force constants to cancel sum-rule violations of the long-range model. In line with our previous work~\cite{Hellman.2013, Knoop.2024}, we impose space-group symmetry and hermiticity, as well as global translational, rotational, and Huang invariances. This way, we can use the long-range model by Gonze and coworkers for crystals of arbitrary symmetry in the way discussed in Ref.\,\cite{Benshalom.2022} while satisfying all of the aforementioned physical constraints.

We note in passing that the non-analytical dynamical matrix in Eq.\,\eqref{eq:dyn.mat.na} is valid for small but finite $\q$ with $q c /\omega \gg 1$ for which the electrostatic approximation holds~\cite{Venkataraman.1975}. This is typically satisfied in Raman backscattering where $\lvert \q \vert \approx 2 k_{\rm i}$. However, for small-angle forward scattering, the resulting $\lvert \q \vert$ can become so small that retardation effects become noticeable, leading to a strongly dispersive modification of Eq.\,\eqref{eq:dyn.mat.na} that restores the analytic behavior of the dynamical matrix. The resulting coupled photon-phonon modes are the so-called polaritons~\cite{Fano.1956, Hopfield.1958}.

\subsection{Evaluation of susceptibility derivatives \label{sec:chi.derivative}}
To evaluate the susceptibility derivatives entering the Raman scattering cross section via Eq.\,\eqref{eq:chi.qs}, we first compute the real-space derivatives $\chi_i^\gamma$ via central differences by displacing each atom in the unit cell by $\pm 0.01$\,\AA{} and computing the susceptibility in the cell via \gls{dfpt}. To obtain the susceptibility derivatives in mode space, $\chi_\qs$, the transformation is performed subsequently using the phonon eigenvectors $\viqs$. This means that for any input $\q$, the same set of susceptibility calculations can be reused, while the spectral function $\Jqs (\w)$ needs to be recomputed. This reduces the number of \gls{dfpt} calculations to $6N$, where $N$ is the number of atoms in the unit cell.
We have verified for \gls{gan} that the resulting activities are insensitive to the size of the simulation cell, i.\,e., larger supercells yield the same Raman activities~\cite{zenodo}. This justifies the averaging scheme leading to the definition of a susceptibility per unit cell entering Eq.\,\eqref{eq:p_iL(t)}.

In the present scheme, two effects are neglected: i) temperature renormalization of the mode activity defined in Eq.\,\eqref{eq:chi.qs} is not accounted for~\cite{Monacelli.2021xgu,Benshalom.2022}, and ii) the electro-optic effect is neglected, where the mode activity of an \gls{lo} mode is changed by the coupling of its macroscopic field to the second-order nonlinear optical coeffcient~\cite{Johnston.1970,Veithen.2005}.

\subsection{Experimental details \label{sec:experiment}}
All measurements were performed in back-scattering geometry on a home-built Raman system based on a 1~m long Horiba FHR-1000 dispersive spectrometer with an 1800 gr/mm holographic grating and a Synapse Plus CCD detector (Horiba Inc.). 
Spectral resolution was $<$0.52~cm$^{\text{-}1}$.
We used Ondax notch filters to allow access to the low-frequency region ($>$10~cm$^{\text{-}1}$) and simultaneous acquisition of the Stokes and anti-Stokes signal.
Undoped polished \gls{gan} single crystals with (0001) and (10$\bar{1}$0) orientations were obtained from MTI Corp.
\gls{bzs} crystals with dimensions of approximately 100 µm were grown using a flux method as reported previously~\cite{Niu.2017}.
\textit{Polarization \gls{gan} measurements:}
We used a 2.54~eV (488~nm) solid-state (Coherent Sapphire SF 488-100 CDRH) laser in ambient conditions with 60~mW excitation power.
To control the polarization of the incident and scattered light (5$^{\circ}$ steps), rotating half-wave plates and a polarizer-analyzer combination were used (see appendix A in \cite{Benshalom.2022} for details).
\textit{Temperature dependent \gls{gan} measurements:}
A 2.33~eV (532~nm) ultra narrow-band frequency-doubled Nd:YAG laser was used (Coherent Prometheus 100) with 22~mW excitation power.
The temperature was set and controlled by a liquid-N$_2$-cooled cryostat (ST-500, by Janis Inc.) and a temperature controller (Lakeshore, Model 335), allowing 30 mins of thermalization at the target temperature between measurements.
\textit{\gls{bzs} measurements:}
We used a 1.58~eV (785~nm) pump-diode laser (Toptica Inc., USA), which is below the band gap of \gls{bzs} (1.9~eV) with 24 mW excitation power.
The sample was mounted into a liquid-He-cooled cryostat, cooled to 10~K at a rate of 1~K/min, and was allowed to thermalize for a few hours prior to measurement.

\subsection{Computational details \label{sec:computations}}
\gls{dft} computations were performed using the \gls{pbe} functional implemented in \aims{} for \gls{gan} \cite{Perdew.1996, Blum.2009, Havu.2009, Knuth.2015}, and the am05 functional for \gls{bzs}~\cite{Armiento.2005}.

For development and production, a \sokrates{} neural network potential was trained on sTDEP samples for \gls{gan} for -1\,\% to +3\,\% strain in a temperature range up to 900\,K, similar to previous work \cite{Frank.2022, Langer.2023ga, Frank.2024}. On an independent test set, the \gls{rmse} of the forces for this potential is 4.1\,{meV/\AA}, which corresponds to a relative error of about 0.4\,\% when normalizing with the \gls{std} of the forces. For low temperatures corresponding to smaller displacements, the errors are accordingly smaller.

\gls{dfpt} calculations were performed with \aims{}~\cite{Blum.2009, Shang.2018} using the PBE functional~\cite{Perdew.1996}. For comparison, further calculations have been performed with \qe{}~\cite{Lazzeri.2003, Giannozzi.2009, Giannozzi.2017, Giannozzi.2020} using Schlipf-Gygi norm-conserving pseudopotentials created with the ONCVPSP code~\cite{Hamann.2013, Schlipf.2015ake}. In that case the \gls{lda} for xc treatment~\cite{Ceperley.1980, Perdew.1981} was used.

The force constants, phonon lifetimes, and spectral functions were obtained using the \gls{stdep} method as described in Sec.\,\ref{sec:stdep} and implemented in the \tdep{} code \cite{Knoop.2024}. Further details are given in the supplemental data~\cite{zenodo}. In the present case, we found it sufficient to neglect the effect of temperature renormalization of the force constants and use temperature-independent force constants obtained at 0\,K using the zero-point motion of the atoms only~\cite{Shulumba.20179s8e,Laniel.2022}. Furthermore, the effect of thermal changes of the internal reference positions, as well as lattice expansion has been found to be negligible for the materials and thermodynamic conditions studied below.

\begin{figure*}[th]
	\centering
	\includegraphics[width=0.9\textwidth]{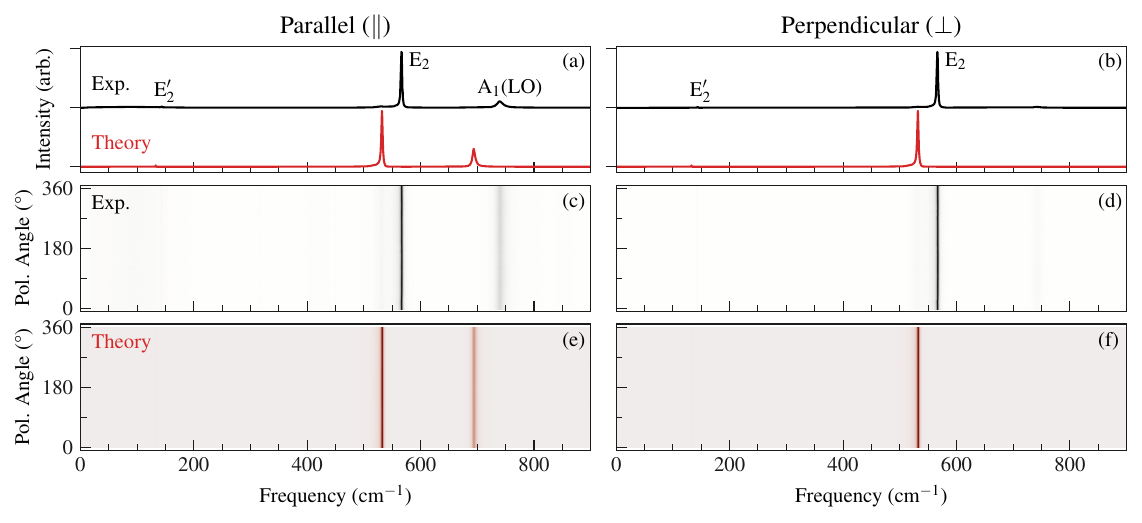}
	\caption{PO Raman results for GaN in (0001) orientation at ambient conditions. (a,b): Unpolarized spectra from experiment (upper black curve) and computation (lower red curve). Curves are normalized, and the upper curve is shifted for clarity. (c-f): Measured/computed polarized scattering intensity as a function of frequency and polarization angle $\poangle$. The intensity scale is linear in all panels.
		\label{fig:gan.po.001}
	}
\end{figure*}

\section{Results \label{sec:results}}

We apply the theory developed in the previous sections to two exemplary materials, wurtzite \gls{gan} and the perovskite \gls{bzs}, and compare the results to experiments performed in this work. \gls{gan} serves as a benchmark material thanks to its simple crystal structure, commercial availability of high-quality samples for experiments, and an accordingly large body of published literature both from experimental and theoretical points of view.
The chalcogenide perovskite \gls{bzs} has recently been suggested as a potential lead-free and thermally stable alternative to halide perovskites for optoelectronic applications, and is under active investigation~\cite{Yuan.2024rqs, Ye.2024, Sopiha.2022gzh, Choi.2022, Nishigaki.2020, Jaramillo.2019, Gross.2017, Wei.202094i}. From a computational point of view, it is much more challenging to simulate than \gls{gan} due to its complex crystal structure and low space-group symmetry, with 20 atoms in the unit cell in the stable \emph{Pnma} phase.

\subsection{PO Raman for GaN}

\begin{figure*}
	\centering
	\includegraphics[width=0.9\textwidth]{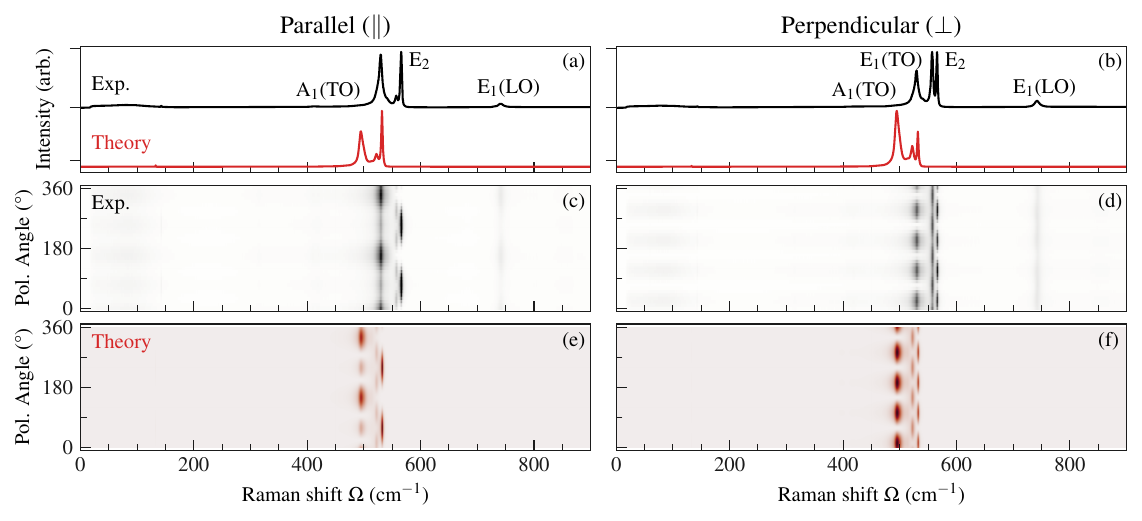}
	\caption{PO Raman results for GaN in ($10\bar{1}0$) orientation at ambient conditions. (a,b): Unpolarized spectra from experiment (upper black curve) and computation (lower red curve). Curves are normalized, and the upper curve is shifted for clarity. (c-f): Measured/computed polarized scattering intensity as a function of frequency and polarization angle $\poangle$. The intensity scale is linear in all panels.
		\label{fig:gan.po.100}
	}
\end{figure*}

Experimental and computed PO Raman maps for the Stokes part of the spectrum at ambient conditions in two different orientations, $\oooi$, and $\ioio$, are presented in Fig.\,\ref{fig:gan.po.001} and Fig.\,\ref{fig:gan.po.100}. Peak frequencies are presented in Tab.\,\ref{tab:gan.frequencies} compared to literature values. The mode assignment follows Ref.\,\cite{Siegle.1995n1b}.
\begin{table}[ht]
	\centering
	\begin{tabular}{lcccccc}
		\toprule
		Reference      & \eiip{} & \aito{} & \eito{} & \eii{} & \ailo{} & \eilo{} \\
		\midrule
		This work      &
		144            & 530     & 557     & 566     & 740    & 742               \\
		Siegle et al.  &
		--             & 533     & 561     & 570     & 735    & 742               \\
		Davydov et al. &
		144            & 534     & 560     & 569     & 737    & 744               \\
		Azuhata et al. &
		144            & 533     & 561     & 569     & 735    & 743               \\
		\bottomrule
	\end{tabular}
	\caption{Experimental mode frequencies for \gls{gan} at room temperature compared to available data from the literature.
		\cite{Siegle.1995n1b,Davydov.1998,Azuhata.19959na}.
	}
	\label{tab:gan.frequencies}
\end{table}

In $\oooi$ direction, the theory reproduces the two prominent peaks, {\ailo} at 740\,\icm, the TO mode \eii{} at 566\,\icm, and a secondary TO mode \eiip{} at 144\,\icm\ with very low intensity, barely visible on the linear scale of the plot. The frequencies are systematically underestimated by about 7\,\%, which is typical for \gls{gga}-level \gls{xc} functionals in \gls{dft}~\cite{Alecu.2010}. We discuss this point further below in Sec.\,\ref{sec:gan.lineshifts}. The PO patterns show no polarization dependence, which is a sign of phase purity of the hexagonal sample since cubic minority phases would produce an angular-dependent pattern for the \ailo{} mode~\cite{Siegle.1995n1b}.

In $\ioio$ direction, the theory reproduces the three prominent peaks from the experiment, i.e., \aito{} at 530\,\icm{}, \eito{} at 557\,\icm, \eii{} at 566\,\icm. In this orientation, each PO pattern has a distinct angular dependence with two to four intensity maxima per full 360$^\circ$ rotation of the polarization directions. As before, the frequencies are systematically underestimated. The faint \eilo{} mode at 742\,\icm is absent in the computation, as it is due to the electro-optic effect, neglected in the current calculation~\cite{Johnston.1970, Cardona.1982, Siegle.1995n1b,Veithen.2003}. All other modes can be assigned.

While not being a focus of this work, we compare the mode activities obtained in this work to those obtained via \qe{}~\cite{Lazzeri.2003, Giannozzi.2009, Giannozzi.2017, Giannozzi.2020}. Therefore, we list isotropic averages according to Eq.\,\eqref{eq:crosssection.iso}
of the mode activities computed on the level of \gls{lda} \gls{dft} as described in Sec.\,\ref{sec:chi.derivative}
in Tab.\,\ref{tab:gan.activity}.
\begin{table}[ht]
	\centering
	\begin{tabularx}{\columnwidth}{X>{\centering\arraybackslash}X>{\centering\arraybackslash}X>{\centering\arraybackslash}X>{\centering\arraybackslash}X}
		\toprule
		Reference          & \eiip{} & \ai{} & \modeEi{} & \eii{} \\
		\midrule
		This work          &
		0.13               & 33.9    & 5.4   & 22.1               \\
		\mbox{QE dynmat.x} &
		0.14               & 34.4    & 6.4   & 23.8               \\
		\bottomrule
	\end{tabularx}
	\caption{Computed mode activities according to Eq.\,\eqref{eq:chi.qs} with isotropic averaging as given in Eq.\,\eqref{eq:crosssection.iso}~\cite{Porezag.1996}, in comparison to Quantum Espresso dynmat.x, with \gls{lda} treatment of \gls{xc}.
	}
	\label{tab:gan.activity}
\end{table}

\subsection{Frequencies and Linewidths for GaN \label{sec:gan.lineshifts}}

\begin{figure}[h!t]
	\centering
	\includegraphics[width=\columnwidth]{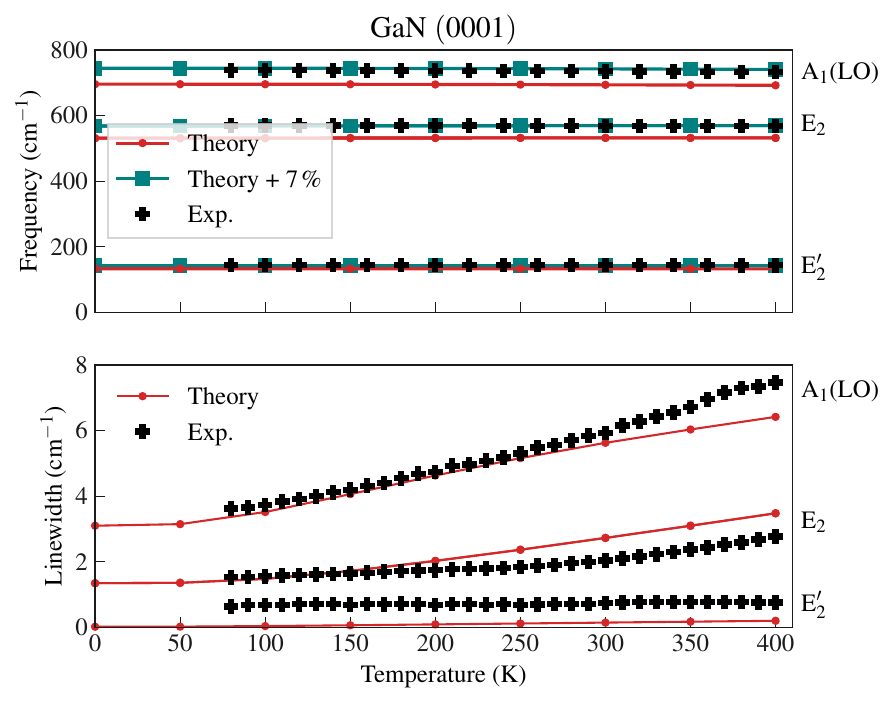}
	\caption{Phonon frequencies and linewidths for Raman active modes in $(0001)$ direction extracted from the experiment (black stars) and computed via perturbation theory (red dots). The theoretical frequencies are systematically underestimated by about 7\,\%, as indicated by the shifted curve (teal squares).
		\label{fig:gan.temperature.001}
	}
\end{figure}

\begin{figure}
	\centering
	\includegraphics[width=\columnwidth]{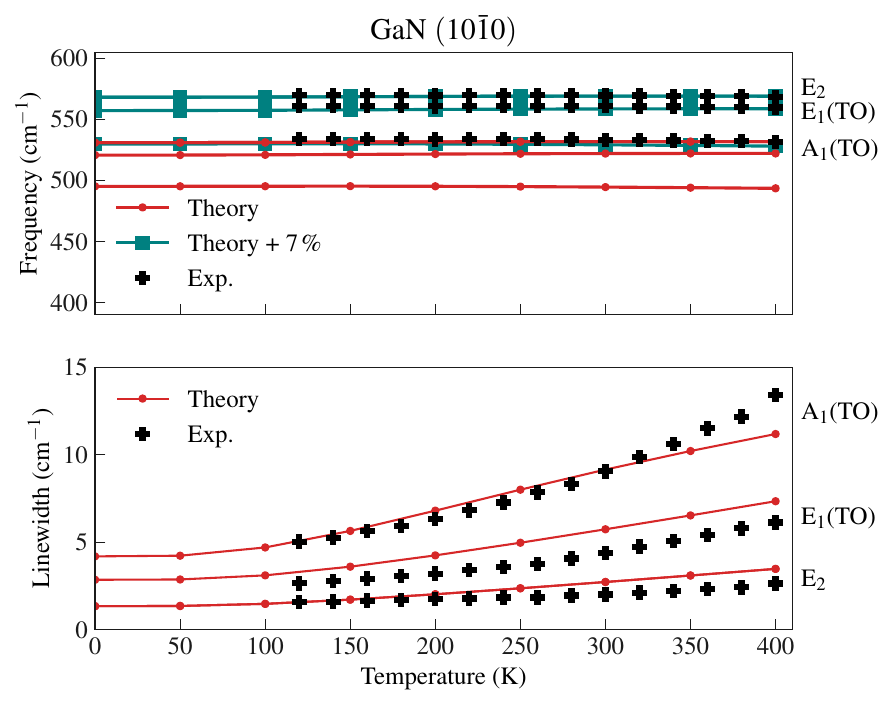}
	\caption{Phonon frequencies and linewidths for Raman active modes in $(10\bar{1}0)$ direction extracted from the experiment (black stars) and computed via perturbation theory (red dots). The theoretical frequencies are systematically underestimates by about 7\,\%, as indicated by the shifted curve (teal squares).
		\label{fig:gan.temperature.100}
	}
\end{figure}

The temperature evolution of frequencies and linewidths for the Raman active modes are shown in Fig.\,\ref{fig:gan.temperature.001} for $\oooi$ orientation, and Fig.\,\ref{fig:gan.temperature.100} for $\ioio$ orientation, both from experiment and computations~\cite{Hellman.2013oi5}. Overall, good agreement is observed for the linewidths of all modes, besides the \eiip{} mode, which is underestimated, and the \ailo{} and \aito{} modes, which deviate above 300\,K.

As mentioned before, the frequencies are systematically underestimated by about 7\,\% (red dots). If the frequencies are corrected by a fixed scaling factor of 7\,\%, the frequencies match very well between experiment and theory across the temperature range, with minimal temperature-induced frequency shifts below 1\,\% in the studied temperature range up to 400\,K.

We note that the effect of LO-TO splitting discussed in detail in Sec.\,\ref{sec:loto} is twofold: First, the frequency differences of 210\,\icm{} between \aito{} and \ailo{}, as well as 183\,\icm{} between \eito{} and \eilo{} modes, are due to LO-TO splitting. Second, the differing phonon dispersions lead to changes in the allowed phonon-phonon scattering events for \gls{to} and \gls{lo} modes, i.\,e., the scattering phase space, which critically influences linewidth broadening in harmonic materials~\cite{Lindsay.2008,Romero.2015}. This explains the different linewidths of \aito{} and \ailo{} modes displayed in Fig.\,\ref{fig:gan.temperature.001} and \ref{fig:gan.temperature.100}.

\subsection{PO Raman for BZS}

As an outlook for more complex materials, we present results for orthorhombic \gls{bzs}, space-group \emph{Pnma}, with 20 atoms in the unit cell. The \gls{bzs} single crystal studied here is the same studied in our earlier work~\cite{Ye.2024}.

\begin{figure*}
	\centering
	\includegraphics[width=0.9\textwidth]{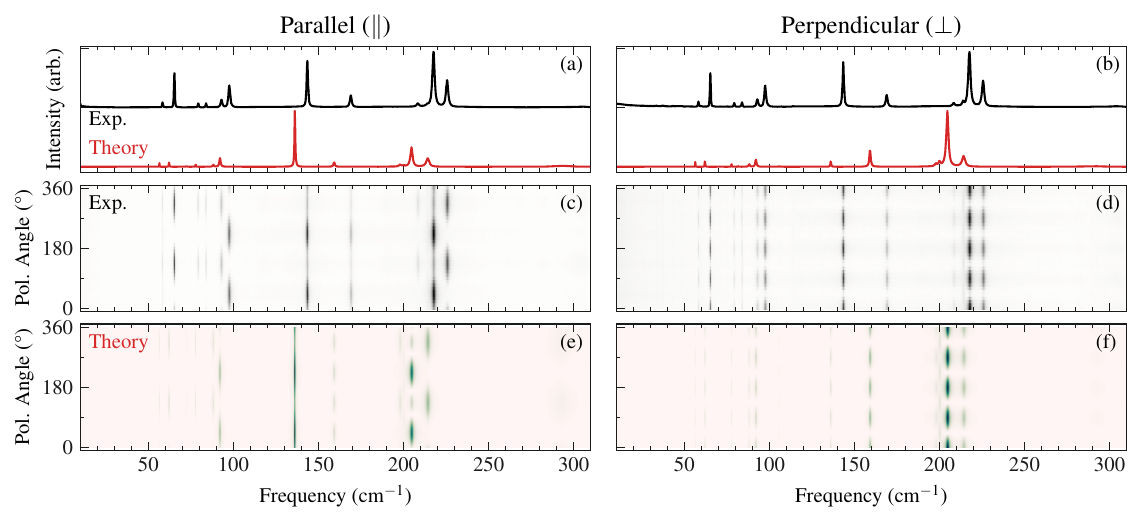}
	\caption{PO Raman results for BZS in (101) orientation at 10\,K. (a,b): Unpolarized spectra from experiment (upper black curve) and computation (lower red curve). Curves are normalized, and the upper curve is shifted for clarity. (c-f): Measured and computed polarized scattering intensity as a function of frequency and polarization angle $\poangle$. The intensity scale is linear in all panels.
		\label{fig:bzs.po}
	}
\end{figure*}

PO Raman maps from experiments at low temperature (10\,K) are presented in Fig.\,\ref{fig:bzs.po}, compared to computed PO maps in $\ioi$ orientation. The orientation was determined by computing PO maps for different crystal orientations and choosing the best available match.
Overall, the match is very good, besides the typical redshift of \gls{gga} \gls{dft} frequencies already observed in \gls{gan}. However, the angular dependence of the PO pattern for each mode matches very well.
Besides the orientation, this again confirms the phase purity of the sample.

\subsection{Unpolarized Raman for BZS at ambient conditions}

To gain more insights into the anharmonic vibrational properties of \gls{bzs}, we further study its Raman response at ambient conditions, where phonon-phonon scattering becomes significant. Experimental and theoretical unpolarized Raman results are presented in Fig.\,\ref{fig:bzs.300K}(a).

The experimental spectrum at 300\,K shows the first-order peaks visible from the low-temperature spectrum, broadened and shifted by anharmonic interactions. Additionally, a broad spectral background appears with some more defined features slightly below 400\,\icm\, potentially from second- and higher-order scattering~\cite{Menahem.2023, Ye.2024}, which we discuss further below.

The computation, according to Eq.\,\eqref{eq:crosssection.1}, reproduces these first-order peaks accurately, hinting at mild anharmonic effects that are well described with perturbative phonon theory~\cite{Hellman.2013oi5, Li.2014}, and confirming thermal stability at ambient conditions as opposed to softer perovskites like lead-halide systems~\cite{Ye.2024}.

\subsection{Simple estimates for second-order Raman scattering in BZS}

Given that the Raman spectrum of \gls{bzs} shows a broad spectral background already at ambient conditions, we discuss some commonly used estimates for second-order Raman scattering based on harmonic phonon theory as presented in Sec.\,\ref{sec:theory.secondorder.raman}.

\begin{figure}
	\centering
	\includegraphics[width=\columnwidth]{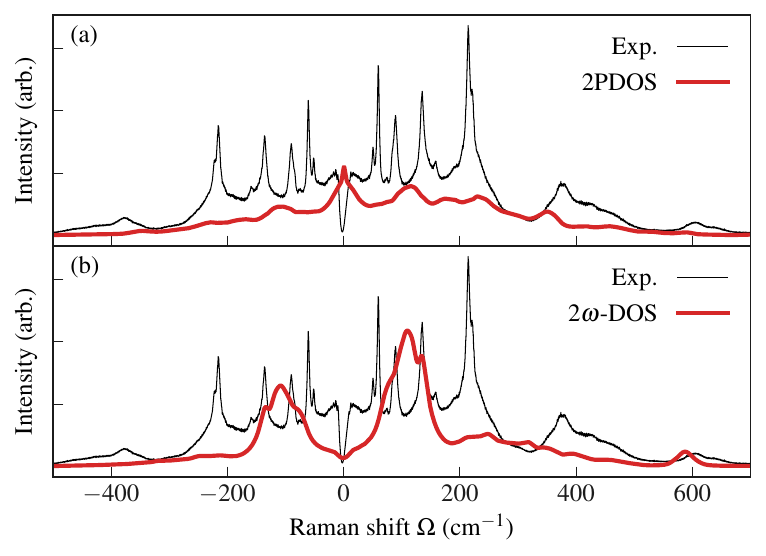}
	\caption{Experimental Raman spectrum for \gls{bzs} at 300\,K (thin black) in comparison to estimated theoretical spectra (red) based on \gls{2pdos} in (a) and the \gls{2wdos} in (b). The intensity of the theoretical spectra is scaled to match the experimental spectrum around 300\,\icm{}.
		\label{fig:bzs.pdos.2wdos}
	}
\end{figure}

We present two popular estimates in Fig.\,\ref{fig:bzs.pdos.2wdos}, the \gls{2pdos} as defined in Eq.\,\eqref{eq:S.2PDOS} and the overtone \gls{2wdos} as defined in Eq.\,\eqref{eq:S.2wDOS}. The intensity of the computed estimates is scaled to match the intensity in the region around 300\,\icm{}. Both estimates give a broad spectrum that is roughly compatible with the experimental one. At the same time, the \gls{2pdos} is smoother and more uniform than the \gls{2wdos} and overall more in line with the experimental spectrum. The \gls{2wdos} predicts a strong spectral weight around 120\,\icm{}, which is not reflected in the experimental spectrum. However, around 600\,\icm\, a peak in the overtone \gls{2wdos} spectrum coincides with an increased response in the experimental spectrum, potentially hinting that this response stems from overtones of the Sulphur-dominated phonon modes around 300\,\icm\ which are barely active in first-order Raman scattering.

\begin{figure}
	\centering
	\includegraphics[width=\columnwidth]{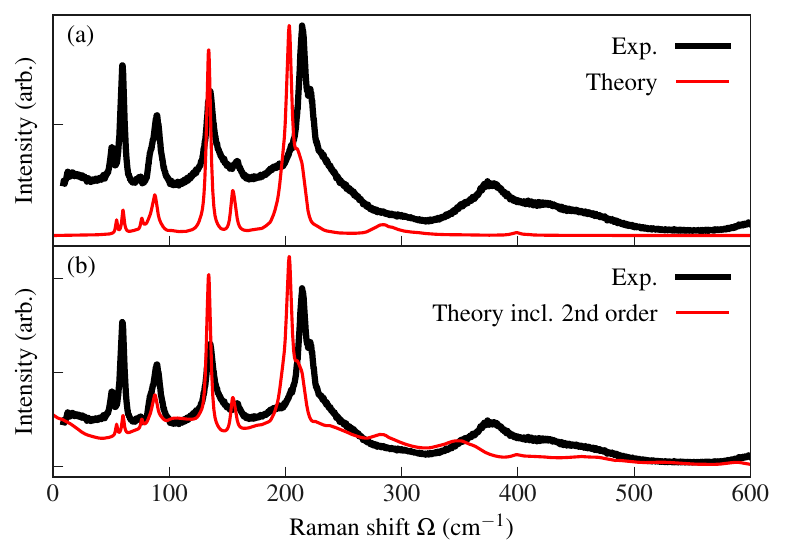}
	\caption{Raman spectrum for BZS at ambient conditions. Upper panel (a): Unpolarized spectra from experiment (black curve) and computation of first-order spectrum (red curve). Lower panel (b): Unpolarized spectra from the experiment (black curve) and computation of first-order spectrum including an estimate of the second-order scattering via autoconvolution of phonon DOS given in Eq.\,\eqref{eq:S.cDOS} (red curve).
		\label{fig:bzs.300K}
	}
\end{figure}

\begin{figure}
	\centering
	\includegraphics[width=\columnwidth]{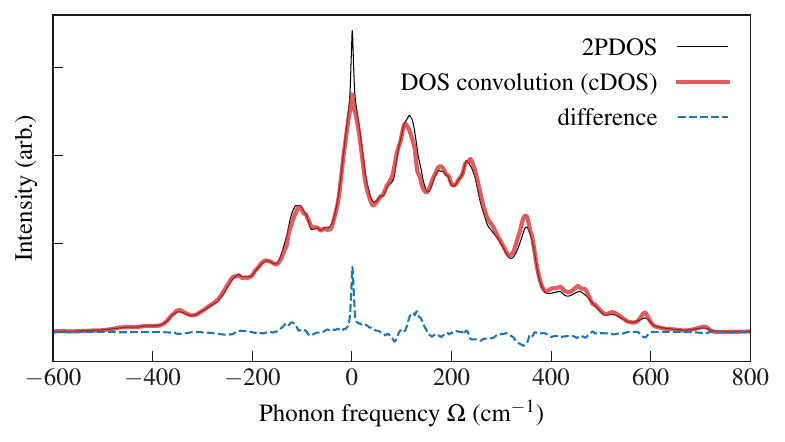}
	\caption{\gls{2pdos} according to Eq\,\eqref{eq:2pdos} (thin black) in comparison to \cDOS (red) according to Eq.\,\eqref{eq:S.cDOS}, and their difference (blue dashed). The two expressions are very similar for \bzs{}.
		\label{fig:bzs.cdos}
	}
\end{figure}

Given that the \gls{2pdos} gives a better estimate for the experimental background spectrum in \gls{bzs}, we calculate another, closely related estimate, from the convolution of the Bose-weighted \gls{dos}, denoted as \cDOS{} in Eq.\,\eqref{eq:S.cDOS}. The comparison between \TPDOS{} and \cDOS{} in \bzs\ is shown in Fig.\,\ref{fig:bzs.cdos}. Both expressions yield a very similar spectrum for \bzs{}, which means that crystal-momentum-conservation is of little importance for the computation of its \TPDOS{}. This finding can likely be generalized to other complex materials with many phonon branches and reinforces the use of \cDOS{} to estimate second-order Raman scattering used in earlier work, c.\,f., Ref.\,\cite{Menahem.2023}.
Motivated by the similarity of \TPDOS\ and \cDOS\, we use the computationally simpler \cDOS\ in Fig.\,\ref{fig:bzs.300K}(b) to correct the computed first-order spectrum with the background estimation based on the \cDOS\, which leads to overall excellent agreement of the spectra besides the second-order feature around 380\,\icm{}.

\section{Conclusion}

We have presented an \emph{ab initio} theory of the Raman scattering process in crystals at finite temperature and applied it to two exemplary materials, \gls{gan} and \gls{bzs}, in comparison to the experiment. We have shown how to include the scattering geometry dependence, focusing on polarization dependence and the small but finite phonon momentum $\q$, which influences the phonon spectral function in optically anisotropic solids with LO/TO splitting. Furthermore, we have shown that perturbative phonon theory based on \gls{dft} simulations can provide an accurate way to describe the spectral properties of mildly anharmonic materials up to a complexity of 20 atoms in the unit cell in the case of \gls{bzs}, establishing the presented method as a unified framework to treat Raman scattering including anharmonic effects in simple to complex materials on the same theoretical footing and in a numerically efficient way. This will be particularly interesting for materials with strong LO/TO splitting or appreciable nuclear quantum effects, for which \gls{md}-based methods require an additional effort~\cite{Raimbault.2019, Kapil.2023, Marsalek.2017}.

While accessing second-order Raman for complex materials within a phonon framework remains a big challenge~\cite{Benshalom.2022}, we have discussed several common estimates for the broad spectral signatures of second-order Raman response in these systems and shown that \gls{2pdos} and a plain convolution of the Bose-weighted \gls{dos} (\cDOS{}) are very similar in \gls{bzs}, allowing to rapidly estimate the second-order background via the \cDOS{} in line with previous work~\cite{Menahem.2023}.

For complex materials, higher-order Raman scattering can likely be more easily accessed with approaches that rely on \gls{md}, accelerated through machine-learned surrogate models for the interatomic forces and dielectric properties~\cite{Raimbault.2019ctl,Gigli.2022,Kapil.2023,Grumet.2024,Berger.2024bc8,Berger.2024,Xu.2024,Rosander.2024}, in particular when coupled with mode mapping for microscopic analysis~\cite{Raimbault.2019, Rosander.2024}.

\gls{md}-based approaches are also preferred for strongly anharmonic systems that feature effects like defect formation~\cite{Knoop.2020, Knoop.2023}, superionic diffusion~\cite{Brenner.2020}, or dynamic disorder~\cite{Hegner.2024}, which are not well described with a phonon picture assuming static reference positions. For anharmonic systems for which a renormalized phonon picture is adequate, the presented approach, as well as related approaches like the \gls{tdscha}, offer benefits in terms of computational effort and high precision~\cite{Monacelli.2021, Monacelli.2021xgu, Miotto.2024}.

While briefly mentioned, non-backscattering geometries and birefringence have not been further discussed in this work. We leave this discussion to future work.

Another direction not discussed in this work is the inclusion of electronic excitations when using laser energies above the band gap of the material of interest, leading to effects like resonant scattering~\cite{Heyen.1990, Sherman.2002}. While computational approaches that take into account resonance effects exist~\cite{Knoll.1995, Ambrosch-Draxl.2002x9u, Gillet.2013, Gillet.2017}, they typically rely on the harmonic approximation and an otherwise unmodified scattering theory. Reconciling anharmonicity with electronic excitations is an interesting future direction.

\section{Code and Data Availability}

All methods to compute phonon spectral functions are implemented and documented in the \tdep{} package~\cite{Knoop.2024}. Additional pre- and post-processing tools necessary to compute and evaluate dielectric derivatives are implemented in \ase{}, and \tdeptools{}~\cite{Larsen.2017,tdeptools}. Code tutorials are available on the \tdep{} website. Project data, plotting scripts, and further supplemental information are available on Zenodo~\cite{zenodo}. Raw \gls{dft} data is available on the NOMAD repository~\cite{nomad,Scheidgen.2023}.

\section{Acknowledgements}

F.K. and O.H. acknowledge support from the Swedish Research Council (VR) program 2020-04630 and the Swedish e-Science Research Centre (SeRC). The computations were enabled by resources provided by the National Academic Infrastructure for Supercomputing in Sweden (NAISS) at NSC and PDC, partially funded by the Swedish Research Council through Grant Agreement No. 2022-06725.
O.Y. acknowledges funding from the European Research Council starting grant (850041 - ANHARMONIC). We acknowledge support from the MIT-Israel Zuckerman STEM Fund and the Sagol Weizmann-MIT Bridge Program. We acknowledge support from the United States-Israel Binational Science Foundation, grant No. 2020270.
F.K. acknowledges Petter Rosander for fruitful discussions about Raman theory and simulations and Samuel Longo for feedback on the implementation. Furthermore, Prakriti Kayastha, Boyang Zhao, David Hardy, and Rafael Jaramillo are acknowledged for discussions related to \gls{bzs}.

\end{document}